\documentclass[aps,prl,reprint,showpacs]{revtex4-1}
\usepackage[dvipdfm]{graphicx}

\draft 

\usepackage{bm}
\def\vb#1{{\bm#1}}
\def\v#1{\mathbf{#1}}			
\def\r{\v{r}} 					
\def\p{\v{p}} 					
\def\R{\v{R}} 					

\def\mat#1#2#3#4{\left(
\begin{array}{cc} #1 & #2 \\#3& #4\end{array} \right)}
\def\del{\partial}

\def\CE{\mathcal{E}}
\def\CO{\mathcal{O}}

\def\CF{\mathcal{F}}

\def\la{\langle}
\def\ra{\rangle}

\begin{document}


\title{Effects of mechanical rotation on spin currents} 



\author{Mamoru Matsuo$^{1,2}$,
Jun'ichi Ieda$^{2,3}$, Eiji Saitoh$^{2,3,4}$,
and Sadamichi Maekawa$^{2,3}$ }
\affiliation{%
$^{1}$Yukawa Institute for Theoretical Physics,  Kyoto University, Kyoto 606-8502, Japan \\
$^{2}$The Advanced Science Research Center, Japan Atomic Energy Agency, Tokai 319-1195, Japan \\
$^{3}$CREST, Japan Science and Technology Agency, Sanbancho, Tokyo 102-0075, Japan\\
$^{4}$Institute for Materials Research, Tohoku University, Sendai 980-8577, Japan}



\date{\today}

\begin{abstract}
We study the Pauli--Schr\"{o}dinger equation in a uniformly rotating frame of reference to describe a coupling of spins and mechanical rotations. 
The explicit form of the spin-orbit interaction (SOI) with the inertial effects due to the mechanical rotation is presented. 
We derive equations of motion for a wavepacket of electrons in two-dimensional planes subject to the SOI. 
The solution is a superposition of two cyclotron motions with different frequencies and a circular spin current is created by the mechanical rotation.
\end{abstract}


\maketitle 

 \emph{Introduction.}---Recently much attention is paid on the control and generation of spin currents, i.e. the flow of electron spins in the field of spintronics\cite{MaekawaEd2006}. 
Since the spin current is a non-conserved quantity, 
the utilization of spin currents is much more challenging than that of charge currents. 
A central concept of spintronics is the transfer of spin angular momentum based on the angular momentum conservation.
Experimental developments in the last decade have allowed us to exchange the angular momentum among conduction electron spin, local magnetization, and photon polarizations. 
These phenomena give birth to a variety of functions\cite{Suzuki2008}, and have accelerated the development of magnetic random access memoriy (MRAM)\cite{Yakushiji2010}. 

In this context, a remaining form of angular momentum carried by condensed matter systems is the mechanical angular momentum due to uniform rotation of a rigid body. 
Using this mechanical angular momentum in spintronics will permit the mechanical manipulation of spin currents. 
However, effects of mechanical rotation on a spin current have not been demonstrated so far.

In this Letter, we derive the fundamental Hamiltonian with a coupling of spin currents and mechanical rotations from 
the generally covariant Dirac equation. 
The introduction of mechanical rotations involves extending our physical system from an inertial to non-inertial frame. 
The dynamics of spin currents is closely related to the spin-orbit interaction (SOI), 
which results from taking the low energy limit of the Dirac equation. 

Figure 1 illustrates the relation between mechanical rotation, magnetization, and spin current.
The coupling of the magnetization and a spin current has been investigated extensively in terms of spin transfer torque\cite{Slonczewski1996,Berger1996}, spin pumping\cite{Tserkovnyak2002}, and spin motive force\cite{Barnes2007}, i.e., the key technologies of spintronics. 
On the other hand, the coupling of a mechanical torque and the magnetization was studied long time ago. 
In the middle of 1910's, the coupling of mechanical rotations and magnetization was investigated by Barnett\cite{Barnett1915}, Einstein and de Haas\cite{Einstein-deHaas1915}. 
They measured the gyromagnetic ratio and the anomalous $g$-factor of electrons before the establishment of the modern quantum physics.
Recently, several groups have detected the effects of mechanical rotations on nanostructured magnetic systems. 
Mechanical detection of ferromagnetic resonance spectroscopy has been recognized\cite{Rugar1992},
Einstein--de Haas effect, rotation induced by magnetization, is observed in the submicron sized thin NiFe films deposited on a microcantilever\cite{Wallis2006}, and 
the nanomechanical detection of a mechanical torque due to 
spin-flips at the normal/ferromagnetic junction of a suspended nanowire has been reported\cite{Zolfagharkhani2008}. 
There is theoretical work on the effects of a mechanical torque acting on a nanostructured magnetic system\cite{Kovalev2003,Kovalev2007,Bretzel2009,Bauer2010}. 
The Einstein--de Haas effect in Bose--Einstein condensates of atomic gases has been proposed\cite{Kawaguchi2006}.
\begin{figure}[b]
\begin{center}
\includegraphics[scale=0.4]{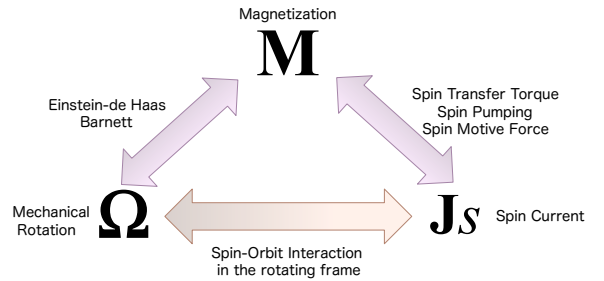}
\end{center}
\caption{Angular momentum transfers between interacting systems. }
\label{motivation}
\end{figure}

Comparing to the well established coupling of mechanical rotations and magnetization, and that of magnetization and spin currents, 
the direct coupling of mechanical rotations and spin currents has not been demonstrated. 
The main purpose of this Letter is to link the mechanical rotation with spin currents.

First, we introduce the Dirac equation in a uniformly rotating frame. 
In the low energy limit of this equation, 
we derive up to the order of $1/m^{2}$ with electron mass $m$ the Pauli--Schr\"odinger Hamiltonian for a single electron including a SOI modified by a mechanical rotation. 
This coincides with that derived by de Oliveira and Tiomno\cite{Oliveira1962} 
when the weak gravitational field in their Hamiltonian is replaced with a uniformly rotating field, in the case of the uniform rotations. 
It is then straightforward to extend the derivation to condensed matter systems in a rotating frame
by replacing the coupling parameter of the SOI in vacuum with that in materials\cite{Nagaosa2006,Chang2008}.
We derive equations of motion for a wavepacket subject to spin-dependent forces due to the SOI term 
and solve them in a particular case. The solution exhibits a circular pure spin current caused by mechanical rotation.

\emph{Dirac equation in a uniformly rotating frame.}---According to the Einstein's principle of equivalence, gravitation cannot be distinguished from noninertiality. 
In the general relativity, both gravitational and inertial effects are expressed by a metric and a connection in a curved space-time.
Dynamics of a spin-1/2 particle in a curved space-time is described by the generally covariant form of the Dirac equation\cite{Nakahara1990}:
\begin{eqnarray}
\left[ \gamma^{\mu} \left( p_{\mu} -  q A_{\mu} - \Gamma_{\mu} \right) + mc/\hbar  \right] \Psi =0,\label{gDirac}
\end{eqnarray}
where $c$ and $\hbar$ are the velocity of light and the Plank constant, $m$ and $q=-e$ are the mass and charge of an electron, 
$ A_{\mu}=(A_{0}, \v{A})$ is the gauge potential, and $ \Gamma_{\mu}$  the spinor connection (see \cite{Nakahara1990}, for example). 
The coordinate-dependent Clifford algebra in the curved space-time $\gamma^{\mu}=\gamma ^{\mu}(x)$ is satisfying $ \{ \gamma^{\mu}(x),\gamma^{\nu}(x) \}=2g^{\mu\nu}(x)$ with the metric $ g^{\mu\nu}(x)$ ($\mu, \nu = 0,1,2,3$). 
In a uniformly rotating frame of reference, of which the angular velocity with respect to an inertial frame is $ \vb{\Omega}(t)$, the coordinate transformation from the rotating frame to the inertial frame is $ d\r' = d\r + \left( \vb{\Omega} \times \r \right)dt$. The space-time line element is given by 
$
ds^{2}= \left[-c^{2} +(\vb{\Omega} \times \r)^{2} \right]dt^{2} + 2(\vb{\Omega} \times \r)dt d\r + d\r^{2}$.
The metric in a uniformly rotating frame becomes
$g_{00}=-1+(\vb{\Omega} \times \r/c)^{2}$, $g_{0i}=g_{i0}=(\vb{\Omega}\times \r/c)_{i}$, 
$g_{ij}=\delta_{ij} \quad (i,j=1,2,3)$.
From this metric, we obtain the Clifford algebra and the spinor connection in the rotating frame as
$\gamma^{0}(x)= i\beta$, $\gamma^{i}(x)=i\beta \alpha_{i}- (\vb{\Omega}\times \r/c)_{i}$,
$\Gamma_{0}= \vb{\Omega}\cdot \vb{\Sigma}/2c, \quad \Gamma_{i}=0$,
where $ \beta=\mat{I}{O}{O}{-I}$ and $ \vb{\alpha}=\mat{O}{\vb{\sigma}}{\vb{\sigma}}{O}$ are the Dirac matrices and $\vb{\Sigma} $ is the spin operator for 4-spinor
defined by $\vb{\Sigma} =\frac{\hbar }{4i} \vb{\alpha} \times \vb{\alpha}$ with  
the Pauli matrix $\vb{\sigma}$. Thus, Eq.(\ref{gDirac}) can be rewritten as
\begin{eqnarray}
&&i\hbar \frac{\del \Psi}{\del t}=H \Psi,  \nonumber \\
&& H = \beta mc^{2} + c \vb{\alpha} \cdot  \vb{\pi} +qA_{0}
	-\vb{\Omega}\cdot \left( \v{r}\times \vb{\pi}+\vb{\Sigma} \right), \label{Hrot}
\end{eqnarray}
where   
$\vb{\pi}=\p - q\v{A} $ is the mechanical momentum and $\r$ is position vector from the rotation axis.
It is well known that, in classical mechanics, the Hamiltonian in the rotating frame has the additional term $ \vb{\Omega} \cdot (\v{r}\times \vb{\pi})$ reproducing the inertial effects: Coriolis, centrifugal, and Euler forces\cite{LandauMech}. 
The term $\vb{\Omega} \cdot \vb{\Sigma}$ is the so-called spin-rotation coupling found in Ref. \cite{Oliveira1962} and also discussed 
 in the context of neutron interferometry in a stationary laboratory on the Earth\cite{Mashhoon1988}. 
The last term of Eq.(\ref{Hrot}), $\v{\Omega \cdot (\v{r}\times \vb{\pi} +\vb{\Sigma})}$, can be regarded as a quantum mechanical generalization of the inertial effects\cite{Mashhoon1988} obtained by replacing the mechanical angular momentum $\v{r}\times \vb{\pi}$ with total angular momentum $\v{r}\times \vb{\pi}+\vb{\Sigma}$. 

\emph{Pauli--Schr\"odinger equation in a rotating frame.}---In the low energy limit, the Dirac equation in a flat space-time reduces to the Pauli--Schr\"odinger equation by the Foldy--Wouthuysen--Tani transformation\cite{Foldy1950,Tani1951}, 
which block diagonalizes the Hamiltonian and is the systematic expansion yielding relativistic corrections 
in any order of the inverse mass, $\CO(1/m^{n}) \, (n=1,2,\cdots)$.
We divide the Hamiltonian (\ref{Hrot}) into even and odd parts 
denoted by $\CE$ and $\CO $, respectively;
$
H=\beta mc^{2} + \CE + \CO, 
\CE=qA_{0}-\vb{\Omega}\cdot(\v{r}\times \vb{\pi}+\vb{\Sigma}), \CO=c {\bm \alpha} \cdot (\p - q\v{A})$.
By successive transformations, the Hamiltonian up to the order of $ 1/m^{2}$ becomes
\begin{eqnarray}
H =&& \beta \left[ mc^{2}+ \frac{\CO^{2} }{2mc^{2}} - \frac{\CO^{4}}{8m^{3}c^{6}} \right] \nonumber\\ 
 &&+\CE - \frac{1}{8m^{2}c^{4}} \left[\CO, \left[ \CO, \CE \right]+ i\hbar \dot{\CO} \right]. \label{Hfwt}
\end{eqnarray}
Neglecting the rest energy in Eq. (\ref{Hfwt}), the Pauli--Schr\"odinger equation for the upper component of Dirac spinors in the rotating frame is obtained by
\begin{eqnarray}
&& i\hbar \frac{\del \psi}{\del t}  = H_{PR} \psi, \label{PReq} \\
&&H_{PR}=H_{K}+H_{Z}+H_{I}+H_{S}+H_{D}, \label{Hpr} \\ 
&&H_{K}=  \frac{1}{2m}  \vb{\pi}^{2} + qA_{0}, \\
&&H_{Z}= \mu_{B} \vb{\sigma} \cdot \v{B} ,\\
&&H_{I}=- \vb{\Omega}\cdot (\r \times \vb{\pi}+\v{S}), \label{HI}  \\
&&H_{S}=\frac{\lambda}{2\hbar} \vb{\sigma}\cdot [ \vb{\pi} \times  q\v{E}'-q\v{E}' \times \vb{\pi} ] , \label{HS} \\
&&H_{D}= -\frac{\lambda}{2} \mbox{div} \left[q\v{E}'\right] ,
\end{eqnarray}
with 
$ \mu_{B}=q\hbar/2m$, $\lambda =\hbar^{2}/4m^{2}c^{2}$, $\v{S}=(\hbar/2)\vb{\sigma}$, and
\begin{eqnarray}
\v{E}'= \v{E} + ( \vb{\Omega}\times \r) \times \v{B}.  
\end{eqnarray}
Equation (\ref{PReq}) is the 2-spinor equation for a single electron in the rotating frame and the Hamiltonian is a 2$\times$2 matrix operator. 
In this expansion, the Hamiltonian to the order of $1/m$ is given by $H_{K}+H_{Z}+H_{I}$. 
The spin-independent $H_{K}$ contains the kinetic energy %
 and the potential energy.
The Zeeman energy $H_{Z}$ contains the $g$-factor of the electron equal to 2, which, combined with $H_{K}$, yields the coupling with magnetic fields,
$(q/2m)(\v{r}\times \vb{\pi}+2\v{S})\cdot \v{B}$.
This contrasts with Eq. (\ref{HI}): the mechanical rotation couples to the total angular momentum of the electron $\v{r} \times \vb{\pi} + \v{S}$. 
The inertial effects, i.e., Coriolis, centrifugal, and Euler forces, are reproduced by $H_{I}$ and $\vb{\Omega}\cdot \v{S}$ is the spin-rotation coupling term.
The expansion of the order of $1/m^{2}$ yields $H_{S}$ and $H_{D}$, which are the SOI and Darwin terms with the mechanical rotation, respectively. 
In the absence of the mechanical rotation, $ \vb{\Omega}=0$, 
these terms reduce to the usual SOI and Darwin terms in a flat space-time. 
In the case of $ \vb{\Omega} \neq 0$, 
we find that 
``the electric force'' $q \v{E}$ is modified by an additional term $(\vb{\Omega} \times \r ) \times \v{B}$. 
This can be interpreted as Lorentz boost with the rotating velocity 
$\vb{\Omega} \times \r$. 
The modified SOI term $H_{S}$ is responsible for the mechanical manipulation of the spin current as shown below.


\emph{Renormalization of SOI.}---
In vacuum, the contribution of $H_{S}$ to $H_{I}$ is negligible, provided that the dimensionless spin-orbit coupling parameter $\eta_{SO}=\lambda (mv)^{2}/\hbar^{2}=(v/2c)^{2} \ll 1$.
However, the effect from the SOI can be enhanced in condensed matter systems yielding renormalization of the coupling $\lambda$ with that of materials. 
Using Fermi momentum $\hbar k_{F}$ as $mv$, $\eta_{SO}$ equals to $\tilde{\lambda} k_{F}^{2}$ where $\tilde{\lambda}$ is a enchanced spin-orbit coupling parameter.
The renormalization depends on detailed electronic structures and electron correlations\cite{Guo2009,Gu2010}.
In the case of Pt, the dimensionless coupling $\eta_{SO}$ is estimated as $0.59$ using the nonlocal measurement of the spin Hall effect \cite{Vila2007,Takahashi2008}. 
Electrons in a non-inertial frame cannot distinguish rotational effects $(\vb{\Omega}\times \v{r}) \times \v{B}$ in Eq. (\ref{HS}) from electric fields $\v{E}$. Therefore, the coupling constant of $(\vb{\Omega}\times \v{r}) \times \v{B}$ is renormalized in the same manner as that of  $\v{E}$.
Consequently, the effect due to the SOI in a rotating frame can be sizable effects as shown below in the large SOI systems\cite{Ast2007,Yaji2010,Guo2009,Gu2010}.

\emph{Circular spin current due to mechanical rotation.}---To clarify physical meanings of the Pauli--Schr\"odinger equation in a uniformly rotating frame, we investigate the equations of motion for operator $\r$, $m \ddot{\r}=\vb{\CF}$, where a quantum mechanical ``force'',
$ \vb{\CF}= \left[ m\v{v},H_{PR}\right]/i\hbar +m \partial \v{v}/\partial t$, 
with $ \v{v}=\left[ \r,H_{PR} \right]/i\hbar$.
Ehrenfest's theorem leads to equations of motion for an electron wavepacket by taking the expectation values with a certain Heisenberg state $\left| \psi \right\rangle $\cite{Chang2008}. 
Since the full expression for $\vb{\CF}$ is lengthy, in this Letter we show a particular case: $\v{E}=\v{0}, \v{B}=(0,0,B),  \v{\Omega}=(0,0,\Omega)$, and $B$ a constant.
In this case, 
the in-plane forces $\langle \vb{\CF} \rangle_{\perp}=(\la \CF_{x} \ra, \la \CF_{x} \ra,0)$ are spin-diagonal and we focus on the electron motion in the $xy$-plane.
Up to the order of $\Omega/\omega_{c}$ with $\omega_{c}=qB/m$, the equations of motion for the center of mass of the electron wavepacket is
\begin{eqnarray}
\ddot{\R}_{\pm}+ a_{\pm} \tau_{y} \dot{\R}_{\pm} - b_{\pm} \R_{\pm}=0, \label{EOM2d}
\end{eqnarray}
where $\R_{+}, \R_{-}$ are the wavepacket's position vector of up- and down-spin electron,
$\tau_{y}=\mat{0}{1}{-1}{0}$ with
$a_{\pm}=(1 \pm \kappa) \omega_{c}$, $b_{\pm}=\pm \kappa \omega_{c}^{2}$, $\kappa = \eta_{SO} \times(\hbar \Omega/2 \epsilon_{F} )$.
$\kappa$ is the dimensionless parameter which separates the electron motion into fast and slow modes.  This parameter consists of the dimensionless SOI coupling $\eta_{SO}$ and the ratio of spin-rotation coupling energy $\hbar \Omega$ to  the Fermi energy $\epsilon_{F}=\hbar^{2}k_{F}^{2}/2m$. 
Because of $|\kappa| \ll 1$ and $a_{\pm} \approx \omega_{c}$, we obtain the solution of Eq. (\ref{EOM2d}) as
\begin{eqnarray}
\R_{\pm}(t) =&& e^{\omega_{c}t \tau_{y}} \tau_{y} \R^{(1)}_{\pm}
+ e^{\pm \kappa \omega_{c} t \tau_{y}}\R^{(2)}_{\pm},\label{Sol}
\end{eqnarray}  
where $\R^{(1)}_{\pm}=\dot{\R}_{\pm}(0)/\omega_{c}$ and $\R^{(2)}_{\pm}=\R_{\pm}(0)- \dot{\R}_{\pm}(0)/\omega_{c}$.
The first term corresponds to the rapid cyclotron motion due to the Lorentz force, $q \v{v} \times \v{B}$, with frequency $\omega_{c}$ and radius $|\R^{(1)}_{\pm}|$.
The second term describes the slow circular motion with the velocity $\v{v}_{d}^{\pm}=\pm R  \kappa \omega_{c} \hat{\vb{\phi}} $ where $\hat{\vb{\phi}}$ is the azimuthal unit vector and radius $R=|\R^{(2)}_{\pm}|$, which is caused by spin-dependent central forces due to the SOI and the mechanical rotation. 
Let us consider an initial condition in which $|\R_{+}(0)|=|\R_{-}(0)|$ and $|\dot{\R}_{+}(0)| = |\dot{\R}_{-}(0)|$. 
Though both up- and down-spin electron move on a circle around $z$-axis with radius $R$, 
each propagates in the opposite direction due to spin-dependence of the second term of Eq. (\ref{Sol}) which originates from the SOI with a mechanical rotation. 
This solution shows that the mechanical rotation causes a circular (pure) spin currents in a rotating frame (Fig. 2).  
The spin current  is obtained as $\v{J}_{s}=\sum_{\sigma =\pm} \sigma e n_{\sigma} \v{v}_{d}^{\sigma}=2en R \kappa \omega_{c} \hat{\vb{\phi}}$ with the electron density $n_{\sigma}=n$.
In the case of $B \approx 1$ T, 
$\Omega \approx 1$kHz, $\eta_{SO} \approx 0.59$, $k_{F} \approx 10^{10}$m$^{-1}$, and $R\approx 0.1 \mbox{m}$,
the spin current $|J_{s}|$ becomes about $10^{8} \mbox{A/m}^{2}$. 
This can be investigated using spin detection methods such as non-local spin valves\cite{Yang2008}, the inverse spin Hall effect\cite{Saitoh2006} and the real-time imaging method\cite{Werake2010}.

The generation of the circular spin currents can be interpreted as an analogy of the drift of charged particles in electromagnetic fields. 
The average velocity of motion of a charge in crossed a magnetic field $\v{B}$ and an external force $\v{F}$ is given by the drift velocity $\v{v}_{d}= \v{F} \times \v{B}/qB^{2}$\cite{LandauFields}. 
In our case, $\v{F}$ corresponds to the spin-dependent force $m b_{\pm} \v{R}_{\pm}= \pm m  \kappa \omega_{c}^{2} \v{R}_{\pm}$.
Thus, the spin-dependent drift velocity is $\v{v}_{d}^{\pm}=\pm m  \kappa\omega_{c}^{2} \v{R}_{\pm} \times \v{B}/qB^{2}$, reproducing the previous result obtained from Eq. (\ref{Sol}). 

\begin{figure}[hb]
\begin{center}
\includegraphics[scale=0.35]{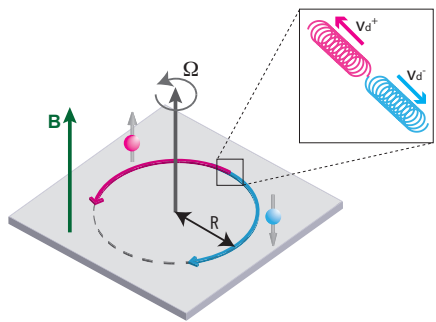}
\end{center}
\caption{Schematic illustration of electrons' trajectories under mechanical rotation $\vb{\Omega}$ and a magnetic field $\v{B}$. 
Solution of equations of motion for wavepacket is a superposition of two cyclotron motions with different frequencies. The drift velocity of the up-(down-) electron is $\v{v}_{d}^{+} (\v{v}_{d}^{-})$ parallel to the azimuthal direction denoted by $\hat{\vb{\phi}}$.
}
\label{circle}
\end{figure}

\emph{Conclusion.}---We have derived the Pauli--Schr\"odinger equation in a uniformly rotating frame of reference thereby describing the coupling of spin to mechanical rotations. 
This equation involves the spin-orbit interaction augmented by a mechanical rotation, 
which reveals a mechanism for the quantum mechanical transfer of angular momentum between a rigid rotation and a spin current.  
Using the semi-classical equations of motion for electrons with spin-dependent forces, 
a circular spin current is predicted. 
It should be noted that starting from the generally covariant Dirac equation is essential when treating spintronics in accelerated frames. 
The present formalism offers a route to ``spin mechatronics'', viz. a strong coupling of mechanical motion with spin and charge transport in nanostructures.




%
%

%

\begin{acknowledgments}
The authors are grateful to G. E. W. Bauer, S. Bretzel, S. E. Barnes, S. Takahashi,  S. Hikino, K. Harii, K. Ando, and J. Suzuki for valuable discussions.  
This work was supported by a Grant-in-Aid for Scientific Research from MEXT, Japan 
and the Next Generation Supercomputer Project, Nanoscience Program from MEXT, Japan.
\end{acknowledgments}


\begin{thebibliography}{10}
\bibitem{MaekawaEd2006}S. Maekawa, ed., {\it Concepts in Spin Electronics} (Oxford University Press, Oxford, 2006).
\bibitem{Suzuki2008}Y. Suzuki and H. Kubota, J. Phys. Soc. Jpn. {\bf 77}, 031002 (2008).
\bibitem{Yakushiji2010}K. Yakushiji \emph{et al.}, Appl. Phys. Express {\bf 3}, 053003 (2010).

\bibitem{Slonczewski1996}J. C. Slonczewski,  J. Magn. Magn. Mater. {\bf 159}, L1 (1996).
\bibitem{Berger1996}L. Berger, Phys. Rev. B {\bf 54}, 9353 (1996).
\bibitem{Tserkovnyak2002}Y. Tserkovnyak, A. Brataas, and G. E. W. Bauer, Phys. Rev. Lett. {\bf 88}, 117601 (2002).
\bibitem{Barnes2007}S. E. Barnes and S. Maekawa, Phys. Rev. Lett. {\bf 98}, 246601 (2007).
\bibitem{Nagaosa2006}N. Nagaosa, J. Phys. Soc. Jpn. {\bf 77}, 031010 (2008).

\bibitem{Barnett1915}
S. J. Barnett, Phys. Rev. {\bf 6}, 239 (1915).
\bibitem{Einstein-deHaas1915}
A. Einstein and W. J. de Haas, Verh. Dtsch. Phys. Ges. {\bf 17}, 152 (1915).
\bibitem{Rugar1992}D. Rugar, C. S. Yannoni, and J. A. Sidles, Nature (London) {\bf 360}, 563 (1992).
\bibitem{Wallis2006}T. M. Wallis, J. Moreland, and P. Kabos, Appl. Phys. Lett. {\bf 89}, 122502 (2006).

\bibitem{Zolfagharkhani2008}
G. Zolfagharkhani \emph{et al.}, Nat. Nanotechnol. {\bf 3}, 720 (2008).

\bibitem{Kovalev2003}A. A. Kovalev, G. E. W. Bauer, and A. Brataas, Appl. Phys. Lett. {\bf 83}, 1584 (2003).
\bibitem{Kovalev2007}A. A. Kovalev, G. E. W. Bauer, and A. Brataas, Phys. Rev. {\bf B 75}, 014430 (2007).
\bibitem{Bretzel2009}S. Bretzel, G. E. W. Bauer, Y. Tserkovnyak, and A. Brataas, Appl. Phys. Lett. {\bf 95}, 122504 (2009).
\bibitem{Bauer2010}G. E. W. Bauer, S. Bretzel, A. Brataas, and Y. Tserkovnyak, Phys. Rev. {\bf B 81}, 024427 (2010).
\bibitem{Kawaguchi2006}Y. Kawaguchi, H. Saito, and M. Ueda, Phys. Rev. Lett. {\bf 96}, 080405 (2006).


\bibitem{Oliveira1962}C. G. de Oliveira and J. Tiomno, Nuovo Cimento {\bf 24}, 672 (1962).
\bibitem{Chang2008}M. Chang and Q. Niu, J. Phys. Condens. Matter {\bf 20}, 193202 (2008).

\bibitem{Nakahara1990}M. Nakahara,  {\it Geometry, Topology and Physics}, (Institute of
Physics Publishing, Bristol, 1998).
\bibitem{LandauMech}L. D. Landau and E. M. Lifshitz, {\it Mechanics}  (Pergamon, Oxford, 1966).
\bibitem{Mashhoon1988} B. Mashhoon, Phys. Rev. Lett. {\bf 61}, 2639 (1988).

\bibitem{Foldy1950}L. L. Foldy and S. A. Wouthuysen, Phys. Rev. {\bf 78}, 29 (1950). 
\bibitem{Tani1951}S. Tani, Prog. Theor. Phys. {\bf 6}, 267 (1951).
\bibitem{Guo2009}G. -Y. Guo, S. Maekawa, and N. Nagaosa, Phys. Rev. Lett. {\bf 102}, 036401 (2009).
\bibitem{Gu2010}B. Gu \emph{et al.}, Phys. Rev. Lett. {\bf 105}, 086401, (2010).
\bibitem{Vila2007}L. Vila, T. Kimura, and Y.C. Otani, Phys. Rev. Lett. {\bf 99}, 226604 (2007).
\bibitem{Takahashi2008}S. Takahashi and S. Maekawa, Sci. Technol. Adv. Mater. {\bf 9}, 014105 (2008).
\bibitem{Ast2007}C. R. Ast \emph{et al.}, Phys. Rev. Lett. {\bf 98}, 186807 (2007).
\bibitem{Yaji2010}K. Yaji \emph{et al.}, Nat. Commun. {\bf 1}, 17 (2010).
\bibitem{Yang2008}T. Yang, T. Kimura and Y. Otani, Nat. Phys. {\bf 4} 851, (2008).
\bibitem{Saitoh2006}E. Saitoh, M. Ueda, H. Miyajima, and G. Tatara. Appl. Phys. Lett. {\bf 88}, 182509 (2006).
\bibitem{Werake2010}L. K. Werake and H. Zhao, Nat. Phys., Advance Online Publication, doi: 10.1038/nphys1742.
\bibitem{LandauFields}L. D. Landau and E. M. Lifshitz, {\it The Classical Theory of Fields}  (Pergamon, Oxford, 1975).


\end{thebibliography}
\end{document}